\documentclass[runningheads,square,comma]{llncs}
\usepackage{graphicx}
\usepackage{booktabs}
\usepackage{multicol}
\usepackage{multirow}
\usepackage{amsmath}
\usepackage{amssymb}
\usepackage{bbding}
\usepackage{subfigure}
\usepackage{colortbl}
\usepackage{pifont}
\usepackage{CJKutf8}
\usepackage{hyperref}       
\usepackage{url}            
\usepackage{amsfonts}       
\usepackage{nicefrac}       
\usepackage{microtype}      
\usepackage{xcolor}         
\usepackage{inconsolata}

\usepackage{bm}
\usepackage{bbm}
\usepackage{gensymb}
\usepackage{color}
\usepackage{mathtools}
\usepackage{algorithm}
\usepackage{algorithmic}

\newcommand\ourmethod{\textsc{LogLG}}

\begin{document}
\begin{CJK*}{UTF8}{gbsn}
\title{\ourmethod{}: Weakly Supervised Log Anomaly Detection via Log-Event Graph Construction}

\titlerunning{LogLG}
\authorrunning{H. Guo et.al.}
%

\author{Hongcheng Guo \inst{1} \and Yuhui Guo \inst{3} \and Renjie Chen \inst{1} \and Jian Yang \inst{1} \and Jiaheng Liu* \inst{1} \and Zhoujun Li* \inst{1} \and Tieqiao Zheng \inst{2} \and Weichao Hou \inst{2} \and Liangfan Zheng \inst{2} \and Bo Zhang \inst{2}}
\institute{Beihang University, Beijing, China \\
\email{\{hongchengguo,crj22,jiaya,liujiaheng,bjq,lizj\}@buaa.edu.cn}
\and Cloudwise Research, Beijing, China \\ 
\email{\{steven.zheng,william.hou,leven.zheng,bowen.zhang\}@cloudwise.com}
\and Renmin University of China, Beijing, China\\
\email{yhguo@ruc.edu.cn}
}


\maketitle
\let\thefootnote\relax\footnotetext{*\quad means the corresponding author.}

\begin{abstract}
Fully supervised log anomaly detection methods suffer the heavy burden of annotating massive unlabeled log data.
Recently, many semi-supervised methods have been proposed to reduce annotation costs with the help of parsed templates. However, these methods consider each keyword independently, which disregards the correlation between keywords and the contextual relationships among log sequences. 
In this paper, we propose a novel weakly supervised log anomaly detection framework, named \ourmethod{}, to explore the semantic connections among keywords from sequences. Specifically, we design an end-to-end iterative process, where the keywords of unlabeled logs are first extracted to construct a log-event graph. Then, we build a subgraph annotator to generate pseudo labels for unlabeled log sequences. To ameliorate the annotation quality, we adopt a self-supervised task to pre-train a subgraph annotator. After that, a detection model is trained with the generated pseudo labels. Conditioned on the classification results, we re-extract the keywords from the log sequences and update the log-event graph for the next iteration.
Experiments on five benchmarks validate the effectiveness of \ourmethod{} for detecting anomalies on unlabeled log data and demonstrate that \ourmethod{}, as the state-of-the-art weakly supervised method, achieves significant performance improvements compared to existing semi-supervised methods.

\end{abstract}

\section{Introduction}

\begin{figure}[t]
	\centering

	\includegraphics[width=0.65\linewidth]{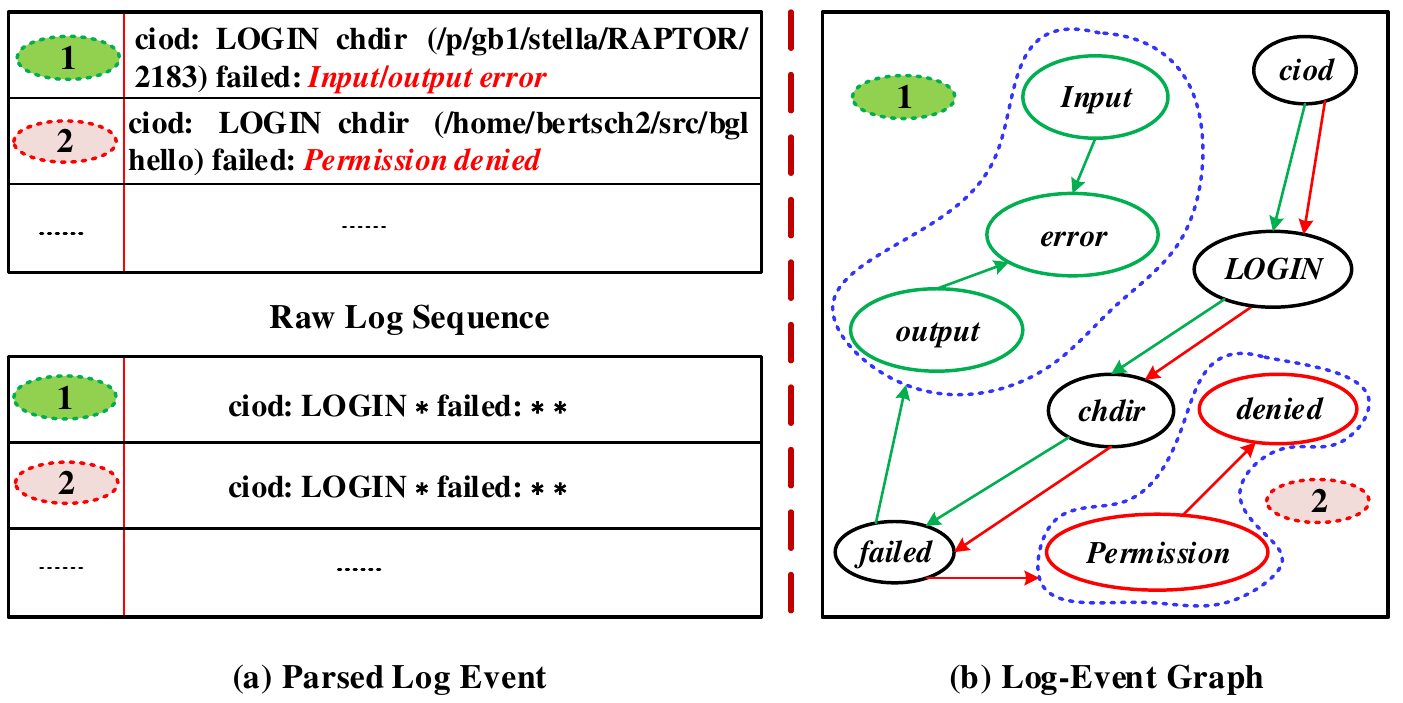}
	\caption{ (a) Number 1 represents log event from normal data while number 2 represents log event from abnormal data. After parsing, they get the same result. (b) Our method exploits the semantic correlation among keywords from different log events by graph. Each event has their corresponding subgraphs, thus we can identify anomaly easily without a parser.}
	\label{intro}
\end{figure}

Due to the recent advancement of the IT industry, software systems have become larger and more complex, which brings huge challenges for maintaining systems~\cite{DBLP:conf/icse/0003HLXZHGXDZ19}. To handle these issues, log analysis \cite{DBLP:conf/sigsoft/ChenYDHZL0ZKGXZ20} is adopted as the main technique to identify faults and capture potential risks. With the development
of deep learning, fully supervised log anomaly detection approaches \cite{DBLP:conf/sigsoft/ZhangXLQZDXYCLC19} have achieved promising successes, which need a large amount of labeled training data consisting of normal and anomalous log sequences. However, labeling massive unlabeled log data is time-consuming and expensive. Thus, semi-supervised learning has attracted increasing interest, resulting in various semi-supervised log anomaly detection methods \cite{DBLP:conf/kbse/LeZ21} are proposed. Such methods generally first employ the log parser to generate templates for labeled normal logs and then detect the anomaly by comparing these templates. So they rely heavily on the log parsing stage\cite{DBLP:conf/icws/HeZZL17}, which means if the log parser does not convert raw logs to the right templates, the detection performance will drop a lot \cite{DBLP:conf/kbse/LeZ21}. Besides, in real industrial scenes, labeled normal data is not easily available due to the evolution of the system. In Figure \ref{intro}, two log events have the same template (\textbf{ciod: LOGIN $*$ failed: $**$}) while one of the two is the anomaly data. This is because valuable tokens (\textbf{Permission denied} and \textbf{Input/output error}) are lost after parsing.

In this paper, to address above shortcomings, motivated by WCKG~\cite{weakly-supervised-text-classification}, we propose a similar weakly supervised framework for \textbf{LOG} anomaly detection via \textbf{L}og-event
\textbf{G}raph construction, called \ourmethod{}. which is an end-to-end iterative method without utilizing both log parsing and labeled normal log data. How can we perform log anomaly detection without massive labeled normal log data? Naturally, we turn our attention to the keywords in the logs at a finer granularity. With the help of Graph Neural Networks \cite{DBLP:conf/iclr/XuHLJ19}, we treat each token as a node in the graph, and by modeling the semantic relationship between nodes, we can obtain the semantics of the whole system. Thus we get rid of the dependence on large-scale normal log sequences. For log parsing, We discard this step, instead, we first sub-word the raw logs and let the model learn to establish connections between important keywords on its own without fixed templates, thus \ourmethod{} is the first end-to-end weakly supervised framework.

Specifically, In each iteration of our model, \ourmethod{} first constructs a log-event graph $G$ with all extracted keywords from log sequences as nodes and each keyword node updates itself via its neighbors.
With $G$, each unlabeled log event $L$ 
corresponds to one subgraph $G_{L}$ of $G$, so we propose a subgraph annotator to assign a pseudo label to each log event $L$. Besides, a self-supervised method is designed to generate more accurate pseudo-labels. Then, we train a classifier to classify all the unlabeled logs with generated pseudo labels.
Based on the classification results, we adopt a simple TF-IDF algorithm to extract more discriminative keywords, where more accurate pseudo labels can be inferred in the next iteration. Extensive experiments on five benchmarks demonstrate that \ourmethod{} effectively detects log anomaly for massive unlabeled log data through a weakly supervised way, and outperforms state-of-the-art methods.

The main contributions of this work are as follows.

\begin{itemize}
	\item We propose a novel weakly supervised log anomaly detection framework, called \ourmethod{}, which 
	explores the semantic correlation among log data via log-event graph without utilizing a parser and massive normal log sequences.
	
	\item An end-to-end iterative process is built to ensure the accuracy of both keywords and pseudo labels, where we design a subgraph annotator to generate pseudo labels for corresponding log events without any labeled log data, thus reducing dependency on log parsing. 
	
	\item We conduct extensive experiments on five benchmark datasets, and the results demonstrate that our \ourmethod{} gains the state-of-the-art performance compared with existing methods. 

\end{itemize}

\section{Related Work}

\subsection{Log Anomaly Detection}

Machine learning based log anomaly detection methods generally contain the following two steps~\cite{DBLP:conf/icdm/LiangZXS07} : (i) Preprocessing log messages, (ii) Anomaly detection. These approaches all require a log parser as a preprocessing operation to extract log templates from log messages. Then, a machine learning model is constructed to detect anomalies. In recent years, deep learning methods have been proposed to improve detection performance \cite{DBLP:conf/ijcai/MengLZZPLCZTSZ19,DBLP:conf/ccs/Du0ZS17,DBLP:conf/icse/YangCWWJDZ21,translog}. Due to the imperfection of log parsing, above methods may lose semantic meaning of log messages, thus leading to inaccurate detection results. Besides, these methods rely heavily on labeled normal data, which is hard to achieve in real world.
\subsection{Pre-trained Model}
\label{Pre-trained models}
BERT~\cite{DBLP:conf/naacl/DevlinCLT19} is a model based on transformer network~\cite{liu2022geometrymotion}, which achieved outstanding performance in various natural language processing tasks~\cite{liu2022cross,guo2022lvp,um4}. One of the major characteristics of BERT is that using two unsupervised learning methods containing masked language modeling (MLM) and next sentence prediction (NSP) to perform the pre-training. The system log can be deemed as sequence data because it is a dataset with an order. In this paper, we use the BERT as the log classifier, and our method directly uses unlabeled raw log data to detect anomalies, which follows an iterative process: generating pseudo labels of log sequences using a subgraph annotator, building a log classifier, and updating the keywords based on the classification results. Among them, the most critical step is generating pseudo-labels for unlabeled raw logs. Then, we employ the raw log with pseudo-labels to train the BERT-based log classifier.

\begin{figure*}[ht]
	\centering
	\includegraphics[width=0.8\textwidth]{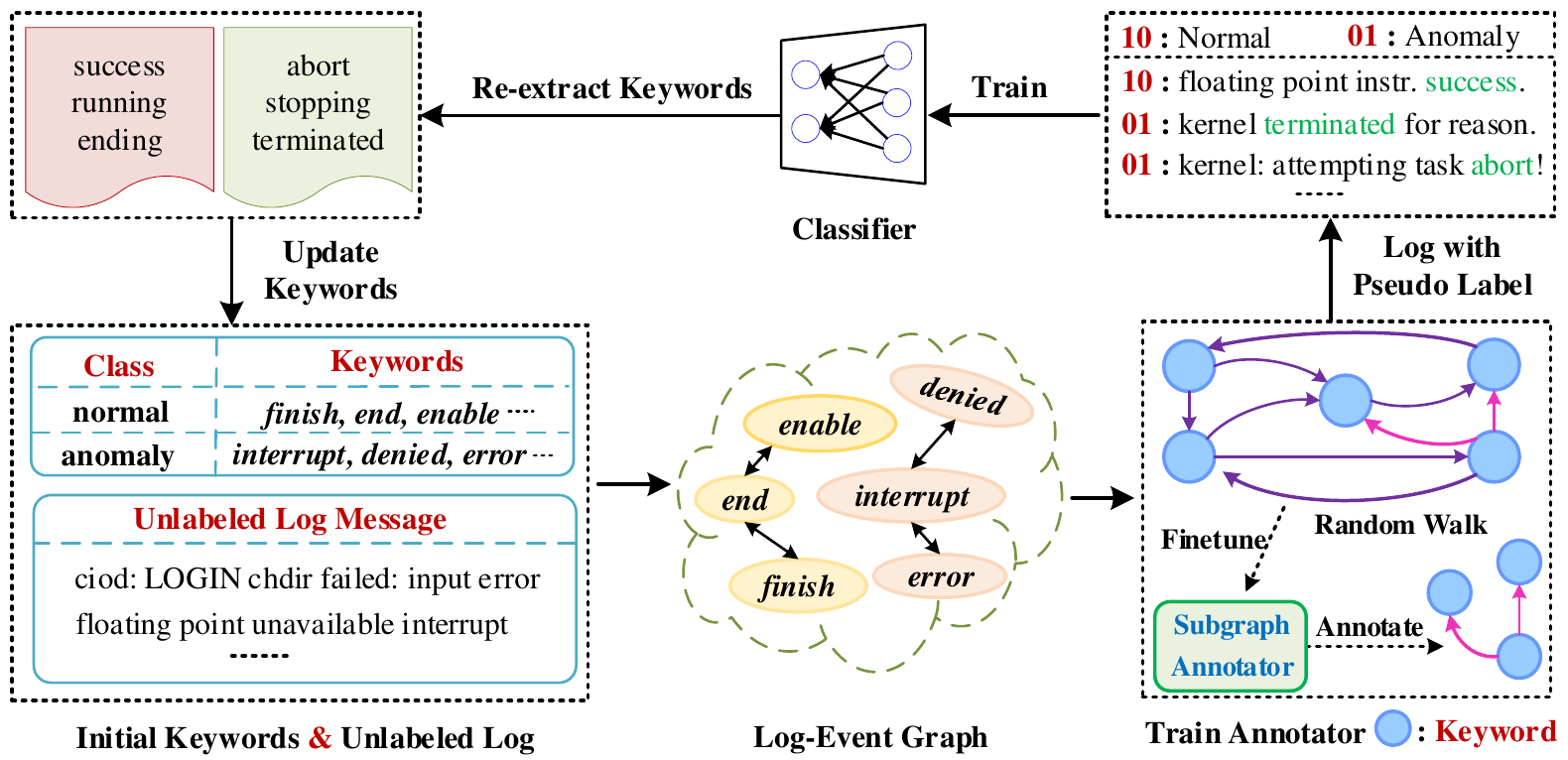}
	\caption{Overview of \ourmethod{}. Our method follows an iterative paradigm. We first build a log-event graph $G$, where unlabeled log sequences correspond to log-event subgraphs of $G$. Then we design a self-supervised task to train the subgraph annotator for pseudo labels, where the random walk is used to generate subgraph as the input of the annotator (i.e., the subgraph formed by pink edges). Then, a log classifier is trained with pseudo labels. Based on the classification, the keywords of log sequences are re-extracted and updated for the next iteration. } 
	\label{overview}
\end{figure*}

\section{Proposed Method}
\subsection{Overview of \ourmethod{}}

In Figure~\ref{overview}, our method is an iterative process, it aims to build a weakly supervised log anomaly detection framework by assigning pseudo labels to unlabeled logs $X$. The input log data of the model contains two parts: (i) A set of initial keywords $L=\{L_{0},L_{1}\}$ extracted by TF-IDF for unlabeled logs, which are randomly initialed into two categories: normal and anomaly, where $L_{i}=\{w_{1}^{i},w_{2}^{i},\cdots ,w_{k_{i}}^{i}\}$ denotes the $k_{i}$ keywords of the category $i$, where $i \in \{0,1\}$. (ii) $n$ unlabeled raw logs $X=\{X_{1},X_{2},\cdots ,X_{n}\}$ from normal and anomaly data.
In each iteration, we first build a log-event graph $G$ based on the keywords from log sequences. To ensure the high-quality of subgraph annotator $A$, we design a self-supervised task to train and fine-tune the annotator $A$ with noisy labels. After that, pseudo-labels are generated by $A$. With the synthetic labels, we train a log classifier. After getting the classification results, keywords $L$ are re-extracted and updated for the next iteration, where the training stage ends when there is no change in the keywords at all.



\subsection{Log-event Graph Construction}

We first exploit the extracted initial keywords of raw log sequences to construct a log-event graph, denoted as $G = (V, E)$, where the keywords as the vertices $V$ and the co-occurrences between keywords as edges $E$. In each iteration, keywords are re-extract by TF-IDF method based on the results of the classifier. The score $S$ of word $w_{m}$ is:

\begin{equation}
	S(w_{m},C_{t})=TF(w_{m},C_{t})\bullet IDF(w_{m})^{M}
\end{equation}

\noindent where $M$ is a hyper-parameter, $C_{t}$ is two classes, $t \in \{0,1\}$. Based on the $S$, the top $Z$ words are selected as the keywords in two classes for the next iteration.
we represent each node $x_{v}$ with the combination of category and index by leveraging the one-hot embedding, thus each node $x_{v} \in R^{2+|V|}$. Then, the semantic correlation knowledge among raw log sequences can be propagated and aggregated by GNN over the log-event graph $G$.

\subsection{Pseudo Label Generation}

After unlabeled log sequences are represented as the subgraphs, the detection of log anomalies is defined as a graph-level classification problem. The keywords of subgraphs belong to a set of vertices in the log-event graph $G$ and the edges among keywords are contained in the log-event graph $G$. Then, we employ Graph Isomorphism Network (GIN) \cite{DBLP:conf/iclr/XuHLJ19} as the subgraph annotator, which is described as:

\begin{equation}
	R_{v}^{(k)}=Liner^{(k)}(\sum_{u\in N(v)}R_{u}^{(k-1)}+(1+\varepsilon ^{k})\cdot R_{v}^{(k-1)})
\end{equation}

\noindent where $R_{v}^{(k)}$ denotes the representation of node $v$ after the $k^{th}$ update. $Liner^{(k)}$ is a multi-layer perceptron in the $k^{th}$ layer. $\varepsilon$ is a learnable parameter. $N(v)$ denotes all the neighbors of node $v$. Then, the generated subgraph $R_{G}$ is represented as follows:

\begin{equation}
	R_{G}=concat[\sum_{v\in G}(R_{v}^{(k)})|k=0 \cdots K]
\end{equation}

\noindent where $\sum_{v\in G}(\cdot)$ denotes the aggregation of all node features from the same layer. GIN concatenates the features from all layers as the subgraph representations.

\subsubsection{Pre-training on Subgraph Annotator}

The pre-training of the annotator is performed on the log-event subgraph. Our self-supervised pre-training process is shown in Algorithm \ref{alg:Framwork}, the subgraph derived from the random walk is similar to the subgraph generated by an unlabeled log sequence, where the subgraph annotator $A$ is trained to predict the class of the start point of a random walk.

\begin{algorithm}[htb] 
	\caption{Pre-training of Annotator on Log-event Graph} 
	\label{alg:Framwork} 
	\begin{algorithmic}[1] 
		\REQUIRE ~~\\ 
		log-event graph $G$, unlabeled log sequences $X$, Gaussian parameters $\mu_{m}$, $\sigma_{m}^{2} $, edge probability $p_{ij}$
		\ENSURE ~~\\ 
		pre-trained subgraph annotator $A$;
		\REPEAT 
		\label{ code:fram:extract }
		\STATE Randomly sample a class $C_{t}$, $t \in \{0,1\}$; 
		\label{code:fram:trainbase}
		\STATE In class $C_{t}$, randomly sample a word $\omega _{t}$; 
		\label{code:fram:add}
		\STATE Sample $L$ from distribution $N (\mu_{m}, \sigma_{m}^{2})$; 
		\label{code:fram:classify}
		\STATE Perform a random walk on $G$, where $\omega _{t}$ is the beginning, $L$ is set to be the length, $p_{ab}$ is the probability, then a subgraph $G_{t}$ is generated; 
		\label{code:fram:select}
		\STATE Input of $A$ is $G_{t}$, $C_{t}$ is the target, loss calculation; 
		\STATE Calculate the gradient and update parameters of $A$; 
		\UNTIL{convergence}
	\end{algorithmic}
\end{algorithm}

First, we randomly sample a keyword $w_{t}$ from one class $C_{t}$. This step is the beginning of a random walk. The process of the random walk follows the Gaussian distribution $N (\mu_{m}, \sigma  _{m}^{2} )$, \textbf{where the number of random walk steps is same as that of the number of keywords contained in an unlabeled log sequence $X_i$.} The parameters of the Gaussian distribution $N(\mu_{m}$, $\sigma_{m}^{2})$ are estimated with $X_{i}$:


\begin{equation}
	\mu_{m}=\frac{1}{n}\sum_{i}^{n}tf(X_{i})
\end{equation}

\begin{equation}
	\sigma_{m}^{2} =\frac{1}{n-1}\sum_{i}^{n}[tf(X_{i})-\mu _{m}]^{2}
\end{equation}

\noindent where $tf(X_{i})$ denotes the number of keywords in a log sequence $X_{i}$. $L$ is the length of random walk from distribution $N (\mu_{m}, \sigma_{m}^{2} )$. Suppose we have two nodes $w_{a}$ and $w_{b}$. The  probability of the random walk from node $w_{a}$ to $w_{b}$ is defined as follows:

\begin{equation}
	p_{ab}=\frac{F_{ab}}{\sum _{\omega _{t}\in N_{\omega _{i}}}F_{at}}
\end{equation}

\noindent where $F_{at}$ is the co-occurrence frequency of $\omega _{a}$ followed by $\omega _{t}$. $N(\omega _{a})$ is the neighbors set of $\omega _{a}$.

Then, we set the node $\omega_{t}$ as the starting node to perform $L$-step random walk. In each step, $p_{ab}$ determines the probability of walking from $\omega _{a}$ to neighbor $\omega _{b}$. Thus when the random walk finishs, a subgraph $G_{t}$ is achieved by us, which is the induced subgraph in the log-event graph $G$. 

In self-supervised pre-training process, we feed the induced subgraph $G_{t}$ into $A$, where the annotator $A$ learns to predict the class of start point $\omega _{t}$. The training objective is defined as the negative log likelihood of $C_{t}$:

Throughout the self-supervised pre-training process, we feed the induced subgraph $G_{t}$ into the annotator $A$, where the annotator $A$ learns to predict the class of the starting point $
\omega _{t}$. Obviously, there are only two categories on this side, and the training objective is defined as the negative log likelihood of $C_{t}$.

\begin{equation}
	L_{self-sup}=-\sum_{r\in rand}C_{t}log(A(G_{t}))
\end{equation}

\subsubsection{Improvement Strategy on Annotator}

After pre-training the subgraph annotator $A$, we design an enhanced strategy to fine-tune it by the voting to generate the labels as follows:

\begin{equation}
	\hat{y}_{a}= arg \underset{k} {max}\{\sum_{b}tf(\omega _{b},X_{a})|\forall (\omega _{b}\in L_{k})\}
\end{equation}

\noindent where $tf(\omega _{b},X_{a})$ denotes the term-frequency (TF) of keyword $\omega _{b}$ in log $X_{a}$. The loss function is defined as:

\begin{equation}
	L_{TF}=-\sum_{a=1}^{n}(kf(X_{i})>0)\hat{y}_{a}log(A(G_{a}))
\end{equation}

\noindent where $G_{a}$ is the subgraph of log sequence $U_{a}$.

\subsection{Training Objective on Log Detection}

After training the subgraph annotator $A$, we use it to generate pseudo labels and annotate all unlabeled logs $X$, which are used to train a log classifier. Our framework can be extended to any log classifier. Generally, the pre-trained BERT \cite{DBLP:conf/naacl/DevlinCLT19} is applied to the fields using sequence data, and the log can be deemed as sequence data. Therefore, we use the BERT as the log classifier. Following the previous works \cite{weakly-supervised-text-classification}, we train the classifier on raw logs with pseudo-labels, the training objective of the classifier is defined as: 

\begin{equation}
	L_{cls} =-\sum_{i=1}^{N} y_{i} \times  log{y}_{i}^{'} 
\end{equation}

\noindent where $y_{i}$ is the one-hot distribution of the real category. Only the probability of the real category is 1, the others are 0. ${y}_{i}^{'}$ is the distribution after $softmax$.

The predicted labels for all unlabeled logs by the log classifier are used to re-extract keywords of log sequences. To decide whether the model has converged, the change of keywords is defined as:

\begin{equation}
	\gamma =\frac{|L^{E_{k}}- L^{E_{k}}\cap L^{E_{k-1}}|}{|L^{E_{k}}|} 
\end{equation}

\noindent where $L^{E_{k}}$ is the keywords set of the $k^{th}$ iteration, $\epsilon$ is a hyper-parameter. When $\gamma<\epsilon$, the iteration finishes.

\section{Experiment}

\subsection{Datasets and Evaluation Metrics}

We validate the effectiveness of \ourmethod{} on five public datasets from LogHub \cite{he2020loghub} \footnote{https://github.com/logpai/loghub}. 
Table \ref{table1} shows the details of datasets. For each dataset, considering that logs evolve over time, we select the first 80\% (according to the timestamps of logs) log sequences for training and the rest 20\% for testing. This setting is the same as the previous work. Since each dataset comes from different systems, thus they require the corresponding preprocessing. We extract the log sequences consistently with the previous works \cite{A2log,DBLP:conf/icse/YangCWWJDZ21}. For \textbf{HDFS}, we generate log sequences according to the $block\_{id}$. For other datasets, we utilize the \textbf{Sliding Window} (size of 20) without overlap to generate log sequences in chronological order. To prevent the model from incorrectly presenting consecutive strings of numbers as keywords, we use a special token \textbf{[Num]} to replace consecutive strings of numbers longer than 4 in each log, while strings of numbers less than or equal to 4 in length are retained. In our experiments, we follow the previous work and adopt evaluation metrics: Precision ($\frac{TP}{TP+FP}$), Recall ($\frac{TP}{TP+FN}$) and $F_1$ Score ($\frac{2*Precision*Recall}{Precision+Recall}$).

\begin{table}[ht]
    \begin{center}
        \caption{A summary of the datasets used in this work. Log Messeages are the raw log strings. Log sequences are extracted by ID or sliding window method.}
        \resizebox{0.6\columnwidth}{!}{
        \begin{tabular}{lcccc}
            \toprule
            Dataset   & Category   &   \#Messages     & \#Anomaly   & \#Templates
            \\
            \midrule
            HDFS   &  Distributed   &    11M  &  17k  &52
            \\
            Hadoop   & Distributed  &  25k    & 368k  & 74
            \\
            OpenStack   & Distributed  &  189k    & 18k & 43
            \\
            BGL    &    Supercomputer     & 5M       & 20k  & 450
            \\
            Thunderbird   & Supercomputer  &  10M    & 123k  & 1108
            \\
            \bottomrule
        \end{tabular}}
        \label{table1}
    \end{center}
\end{table}

\subsection{Implementation Details}

The training and evaluation are performed on
NVIDIA RTX 2080Ti. Subgraph annotator is a three-layer GIN \cite{DBLP:conf/iclr/XuHLJ19}. The training epoch for the annotator is 30. We set the batch size of self-supervision to 20. For the classifier, we set the batch size to 5 for log sequences. Both the subgraph annotator
and the classifier use Adam as optimizer. Their learning rates are 1e-4 and 2e-6, respectively. The classifier uses BERT-base-uncased for initialization. For keyword extraction,
we select top 100 keywords respectively for abnormal and normal data. The hyper-parameter $M$ is set to 4.

\subsection{Comparison with state-of-the-art Methods}

\begin{table*}[!ht]
\centering
\setlength{\tabcolsep}{6pt}
\caption{Performance of our model compared with baselines. \ourmethod{} obtains state-of-the-art results among five datasets.}
\resizebox{0.95\columnwidth}{!}{
\begin{tabular}{c| c |c|c c c c c} 
\toprule
\multirow{2}{*}{\textbf{Model}} 
& \multirow{2}{*}{\textbf{Type}}
& \multirow{2}{*}{\textbf{Log Parser}} & \textbf{HDFS} & \textbf{BGL} & \textbf{Thunderbird} & \textbf{Hadoop} & \textbf{OpenStack} \\
& & & \textbf{$F_1$ Score} & \textbf{$F_1$ Score}  & \textbf{$F_1$ Score} & \textbf{$F_1$ Score} & \textbf{$F_1$ Score} \\ \midrule

\multirow{2}{*}{\textbf{DeepLog~\cite{DBLP:conf/ccs/Du0ZS17}}}
       & {semi-supervised} & \Checkmark & 0.875 & 0.227 & 0.823 & 0.742 & 0.857\\
       &{semi-supervised}  & \XSolidBrush & 0.221 & 0.088 & 0.172 & 0.102 & 0.207\\ \midrule
\multirow{2}{*}{\textbf{LogAnomaly~\cite{DBLP:conf/ijcai/MengLZZPLCZTSZ19}}}                                                                         
    &{semi-supervised} & \Checkmark & 0.878 & 0.303 & 0.843 & 0.759 & 0.851 \\
    & {semi-supervised} & \XSolidBrush & 0.268 & 0.101 & 0.231 & 0.119 & 0.198 \\ 
    \midrule
\textbf{A2log~\cite{A2log}}
    & {semi-supervised} & \Checkmark & - & 0.320 & 0.940 & - & - \\\midrule
\textbf{LAnoBERT~\cite{DBLP:journals/corr/abs-2111-09564}}
    & {semi-supervised} & \XSolidBrush & 0.954 & 0.874 & - & - & - \\\midrule
\multirow{2}{*}{\textbf{PLELog~\cite{DBLP:conf/icse/YangCWWJDZ21}}}
    & semi-supervised & \Checkmark & \textbf{0.957} & \textbf{0.976} & 0.949  & 0.841 & 0.866 \\
    & semi-supervised & \XSolidBrush & 0.264 & 0.223 & 0.213 & 0.247 & 0.195 \\  \midrule
\multirow{1}{*}{\textbf{\ourmethod{}}}
    & weakly supervised & \XSolidBrush & 0.955 & 0.963 & \textbf{0.968}  & \textbf{0.875} & \textbf{0.912} \\ 
\bottomrule
\end{tabular}}
\label{baselines}
\end{table*}

\subsubsection{Performance Comparison}

In Table \ref{baselines}, \ourmethod{} achieves the competitive performance compared with baselines in terms of $F_{1}$ score. Labeled positive data provides more distribution cues for semi-supervised methods to detect anomalies. We find that some baselines perform worse on BGL like DeepLog and LogAnomaly. This is because of the diversity of log events of BGL. Faced with such erratic log data, \ourmethod{} still performs well. Although there are new log events in the test set, most of the keywords in these events have appeared in the training set. \ourmethod{} seeks keyword associations at a finer granularity, which enhances the performance for detecting anomalies. 

Compared with PLElog, \ourmethod{} achieves significant improvements on these benchmark datasets without using the log parser. \ourmethod{} requires no labeled data during the training stage while PLElog relies on massive labeled normal data, which is costly and impractical in industry scenes. Besides, PLElog requires a parser to pre-process the raw log data while \ourmethod{} is an end-to-end framework without a parser. Besides, other baselines perform badly without a parser too, confirming that for template-based methods, the parser stage is heavily needed. For example, Deeplog drops from 0.875/0.823 to 0.221/0.172 on HDFS/Thunderbird.

\subsection{Ablation Study}

In this section, we mainly validate the effects of key components
and parameters in our framework. Experiments are conducted on the Thunderbird.



\subsubsection{\textbf{Effectiveness of Subgraph Annotator}}

To demonstrate the effect of the annotator, we compare the results with/without the subgraph annotator and with/without self-supervision (SS) on Thunderbird. For the case without a subgraph annotator, we determine to generate pseudo-labels through keyword counting, which is widely operated under weak supervision tasks. For the case without self-supervision, we directly fine-tune the subgraph annotator. 

\begin{figure}[ht]
	\centering
	\includegraphics[width=0.65\linewidth]{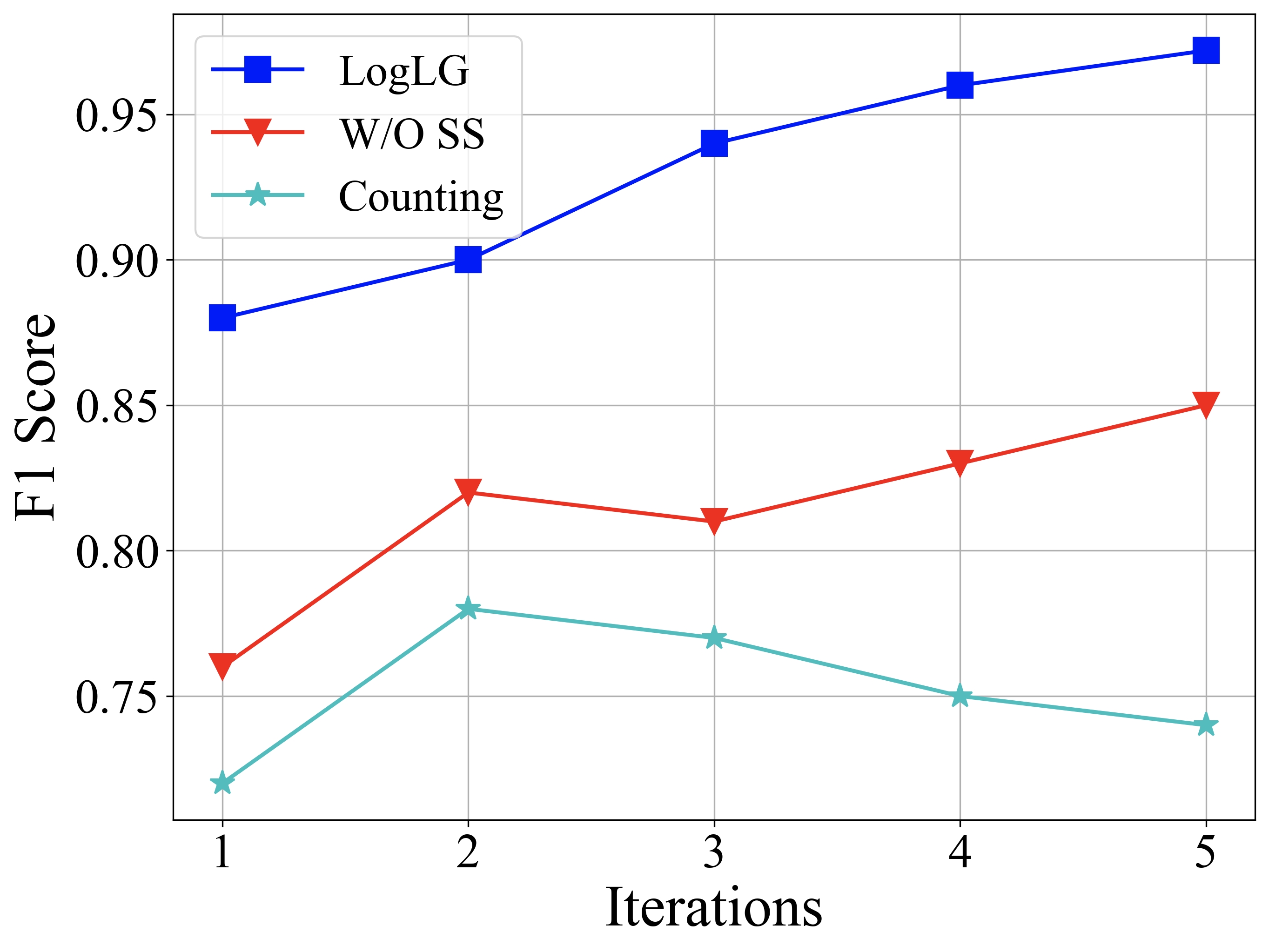}
	\caption{Results on Thunderbird. where \emph{W/O SS} represents the case without self-supervision and \emph{Counting} represents the case leveraging keyword counting method to generate pseudo-labels.}
	\label{anotator}
\end{figure}

In Figure \ref{anotator}, we can see that 1) our method with all components of the annotator obtains the highest performance, proving the effectiveness of our subgraph annotator. 2) For the case using keyword counting, the quality of pseudo labels is the worst, which leads to the worst classification performance. 3) For the case with fine-tuning but no self-supervised learning, it outperforms the keyword counting by 11$\%$ on $F_1$ score. 4) Self-supervised learning task
can boost the performance, exceeding the case without SS by a large margin of 12$\%$ on $F_1$ score.

\begin{figure}[ht] 
	\centering 
	\begin{minipage}[t]{\linewidth} 
		\subfigure{
			\begin{minipage}[t]{0.47\linewidth} 
				\centering
	    	\includegraphics[width=1.6in,height=1.3in]{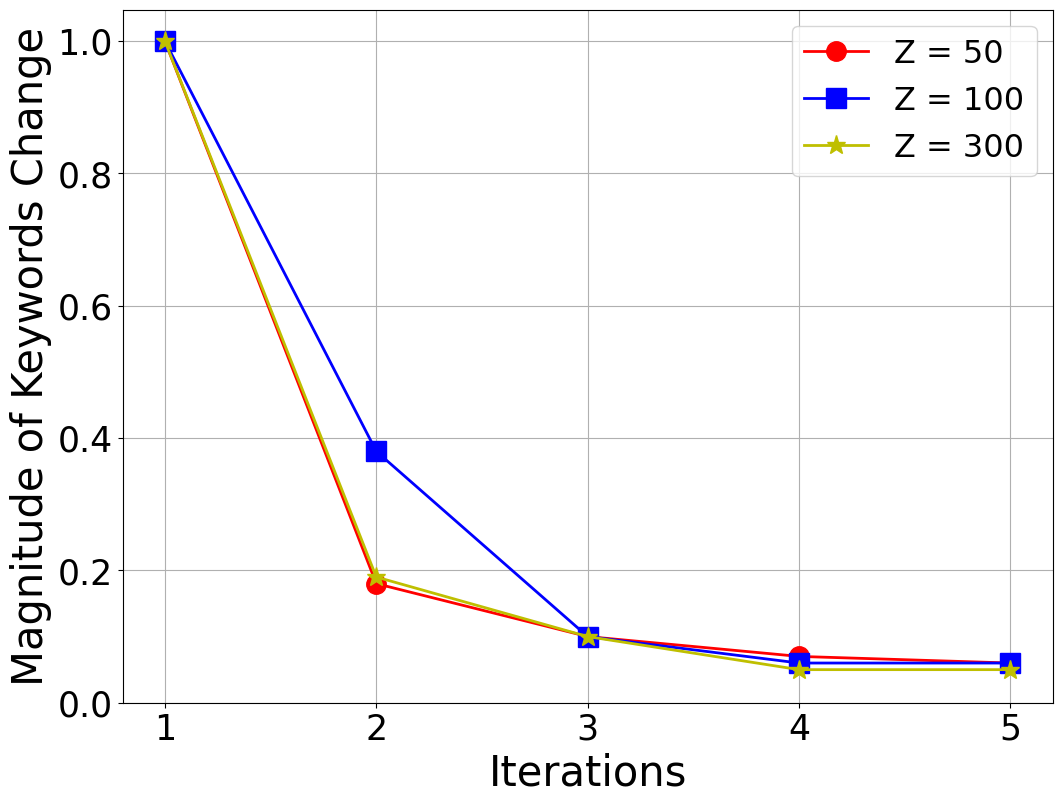}
				
			\end{minipage}
		}
		\subfigure{
			\begin{minipage}[t]{0.47\linewidth} 
			\centering
				
		    \includegraphics[width=1.6in,height=1.3in]{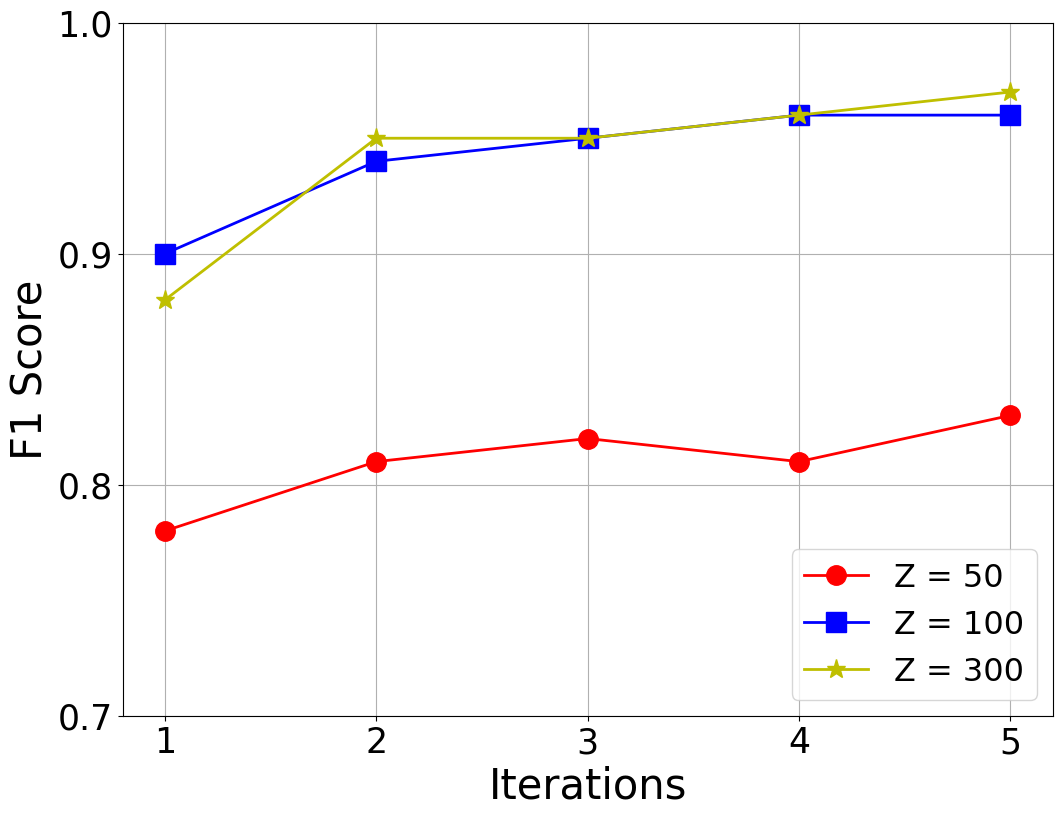}
				
			\end{minipage}
			
		}
	\end{minipage}
	\vfill
	\caption{Effect of keywords number $Z$ on the change of keywords (at left) and the detection results (at right).}
    \label{keyword results}
\end{figure}

\subsubsection{Effect of the Number of Keywords}

Here, we validate the effect of the number $Z$ of extracted keywords in Figure \ref{keyword results}. We can see that (i) The change of keywords falls below 0.1 in the 3rd update for all three number settings. (ii) Increasing the
number of keywords from 50 to 100 brings a significant
performance improvement, while more keywords
($Z$ = 300) make little change.

				
				
				
			



\subsection{Case Study}

Here, we present a case study to show the power of our framework. In the beginning, we
take “failed” as the initial keyword by TF-IDF. After two iterations, the keywords are updated
and the top 12 keywords are presented in Table \ref{kwords}. Obviously, top 12 keywords extracted are correspond to the exceptions, belonging to “Anomaly”.
By comparing the keywords between the 1st and the 2nd, we discover that our method has the ability to seek more accurate keywords during iteration. Besides, since we kept the numbers in messages, we could find some combinations of keywords and numbers in the results. For example, 'infinihost0' represents the first host. After analysing, we are surprised to find that these combinations often represent a specific process or host. Exception is injected and passes through this host, which means our model is equipped with the ability of localizing anomalies. Based on this finding, it is possible for \ourmethod{} to trace and conduct root analysis on anomalies, which is far from what those template-parsing based methods can do. 

\begin{table*}[!ht]
    \centering
    \caption{Top 12 keywords. Top 12 keywords belonging to \emph{Anomaly} class are list out in the first two iteration. We select \emph{``failed''} as the initial keyword by TF-IDF.}
    
    \resizebox{0.9\columnwidth}{!}{
    \begin{tabular}{c c c c c}
        \toprule
        Iteration  &Keywords \\
        
        \midrule
            
        0            & failed   \\
        \midrule
        \multirow{2}{*}{1}  
                    &  denied, failed, ignoring, obj\_host\_amd64\_custom1\_rhel4, error, append \\
                     & errbuf, tavor\_mad, unexpected, get\_fatal\_err\_syndrome, ram0, infinihost0\\
        \midrule
        \multirow{2}{*}{2}             & denied, ignoring, infinihost0, failed, error, errbuf\\
                     & unexpected, null, get\_fatal\_err\_syndrome, unconfined, append, obj\_host\_amd64\_custom1\_rhel4\\
                     
        \bottomrule
    \end{tabular}}
    \label{kwords}
\end{table*}

\section{Conclusion}

In this paper, we propose a novel  end-to-end \ourmethod{} method for
weakly supervised log anomaly detection via
log-event graph construction. In each iteration, we first construct a log-event graph and transform every unlabeled log sequence into a subgraph. To accurately annotate the subgraphs, we first pre-train the subgraph annotator with a designed self-supervised task and then fine-tune it. Under the generated pseudo labels, we train a detection model to classify the unlabeled data. Then we re-extract keywords from log sequences to update the log-event graph. Extensive experiments on benchmarks demonstrate that \ourmethod{} method without log parsing significantly outperforms existing semi-supervised methods for detecting anomalies on unlabeled log data.


\clearpage
\bibliography{main}
\bibliographystyle{splncs04}

\end{CJK*}
\end{document}